\documentclass[lineno]{jfm}

\usepackage{newtxtext} 
\usepackage{newtxmath}

\usepackage{natbib}
\usepackage{hyperref}
\usepackage[nameinlink,capitalise,noabbrev]{cleveref}

\title{On-off pumping for drag reduction in a turbulent channel flow}

\author[G. Foggi Rota, A. Monti, M. E. Rosti and M. Quadrio]
{Giulio Foggi Rota\aff{1,2}, Alessandro Monti\aff{1}, Marco E. Rosti\aff{1} and Maurizio Quadrio\aff{2}}

\affiliation{
\aff{1} Complex Fluids and Flows Unit, Okinawa Institute of Science and Technology Graduate University (OIST), 
1919-1 Tancha, Onna-son, Okinawa 904-0495, Japan
\aff{2} Dipartimento di Scienze e Tecnologie Aerospaziali, Politecnico di Milano,
via La Masa 34, 20156 Milano, Italy
}

\begin{document}

\maketitle

We show that the energy required by a turbulent flow to displace a given amount of fluid through a straight duct in a given time interval can be reduced by modulating in time the pumping power.
The control strategy is hybrid: it is passive, as it requires neither a control system nor control energy, but it manipulates how pumping energy is delivered to the system (as in active techniques) to increase the pumping efficiency.
Our control employs a temporally periodic pumping pattern, where a short and intense acceleration (in which the pumping system is on) followed by a longer deceleration (in which the pumping system is off) makes the flow alternately visit a quasi-laminar and a turbulent state. 
The computational study is for a plane channel flow, and employs direct numerical simulations, which present specific computational challenges, for example the highly varying instantaneous value of the Reynolds number, and the importance of discretisation effects.
Particular care is devoted to a meaningful definition of drag reduction in the present context. 
The ability of the forcing to yield significant savings is demonstrated. Since only a small portion of the parameter space is investigated, the best performance of the control technique remains to be assessed.

\section{Introduction}
\label{sec:intro}

In the ever-growing field of flow control for turbulent drag reduction, 
techniques are conventionally grouped into active and passive, with the 
latter typically including special roughness (e.g. riblets) and other 
boundary treatments for which no control energy is required.
In this paper, concerned with skin-friction drag reduction in a turbulent plane channel flow, we demonstrate the success of an approach that requires no control system or energy as in passive techniques, and delivers pumping energy in a time-varying manner, eventually increasing the pumping efficiency.

Modulating in time the injection of the pumping power into the system is one of the 
least considered (yet one of the simplest) approaches to flow control. 
Recent progresses in understanding the transient nature of 
turbulence point to the potential of an unsteady power delivery, designed to move 
the flow back and forth between the laminar and turbulent regimes.
The strategy studied in this work, and preliminarily introduced by \cite{foggirota-etal-2023}, consists of a periodic pumping with an on-off waveform, which produces a sequence of accelerations and decelerations. 

The literature concerning sudden flow accelerations and decelerations \citep{mathur-etal-2018}, as well as periodic pulsatile flows \citep{xu-song-avila-2021}, is vast. Since the seminal study of \citet{tu-ramaprian-1983}, the ability of the oscillating forcing to alter the mean flow and to achieve drag reduction has been a debated issue.
It has been shown \citep[see for example][]{manna-vacca-verzicco-2015} that a reduction of the mean turbulent friction is possible by harmonically modulating the driving pressure gradient. 
This strategy, however, requires a huge amount of extra energy, and results in an overall decreased efficiency: the total energy would be better spent for conventional, steady pumping.
 
Although most studies employ harmonic pulsations of the streamwise pressure gradient 
superimposed to a steady component, it is also known that non-sinusoidal waveforms
\citep{brindise-vlachos-2018, ciofalo-2022} might produce a different response of the flow, with the deceleration time and the acceleration intensity being the parameters 
transition is most sensitive to. 
For an efficient use of pumping energy, we employ a strategy 
inspired by \cite{iwamoto-sasou-kawamura-2007}, who numerically studied 
a plane channel flow \citep[an experimental followup was presented by][for a pipe flow]{souma-iwamoto-murata-2009} driven by a time-varying pressure gradient $-\Pi(t)$. The periodic function $-\Pi(t)$ was a square wave of period $T$, with average value $-\overline{\Pi}$, cyclically alternating between a negative value $-\overline{\Pi} - \Delta \Pi$ during one half of the cycle and a positive value $-\overline{\Pi} + \Delta \Pi$ (since $\Delta \Pi > \overline{\Pi}$) during the other half.
By judiciously choosing the value of the period, they found that the cycle-averaged 
skin friction can be reduced compared to a canonical channel flow 
that produces the same flow rate. 
Recently, the same group \citep{kobayashi-etal-2021} replicated the experimental 
study for a low-Reynolds pipe flow; the parameter space was explored in depth 
by automatically generating more than $7000$  waveforms producing approximately the same flow rate, and confirming the reduction of the cycle-averaged skin friction.
However, these promising results should be considered together with their limitations. In their original study, the baseline Reynolds number was highly subcritical, and only five periods and one waveform with a single amplitude were investigated; moreover, the changing sign of $\Pi(t)$ during the cycle implies that energy recovery during one half of the cycle is needed to achieve true savings, which renders the strategy technologically complex. 

In this paper, we consider a time-varying pressure gradient that yields large reductions of drag in a turbulent plane channel flow, using high-fidelity direct numerical simulations (DNS).
A single temporal waveform for the pressure gradient, made by a simple on-off sequence, is employed. The energetic efficiency of the procedure and its variation with the control parameters are assessed, and the dynamics of relevant bulk quantities is discussed.

The paper is organised as follows: in \S\ref{sec:setup} we describe 
the problem setup and the nature of the forcing, providing 
details on our numerical methods and procedures. 
We also elaborate on the concept of "drag reduction", 
which becomes particularly delicate in the present context, owing to the lack of a unique reference flow configuration to compare with.
The main results of the numerical study are presented in \S\ref{sec:results}, followed by concluding remarks in \S\ref{sec:conclusions}. 

\section{Problem setup}
\label{sec:setup}

\subsection{The on-off pumping}
The numerical study is carried out by direct numerical simulation of  
the incompressible Navier--Stokes equations in an indefinite plane channel. The geometry consists in two planar walls, separated by a distance $2h$; a Cartesian coordinate system is employed, with axes $x$, $y$ and $z$ aligned with the streamwise, wall-normal and spanwise directions. 

\begin{figure}
\centering
\includegraphics[width=1\textwidth]{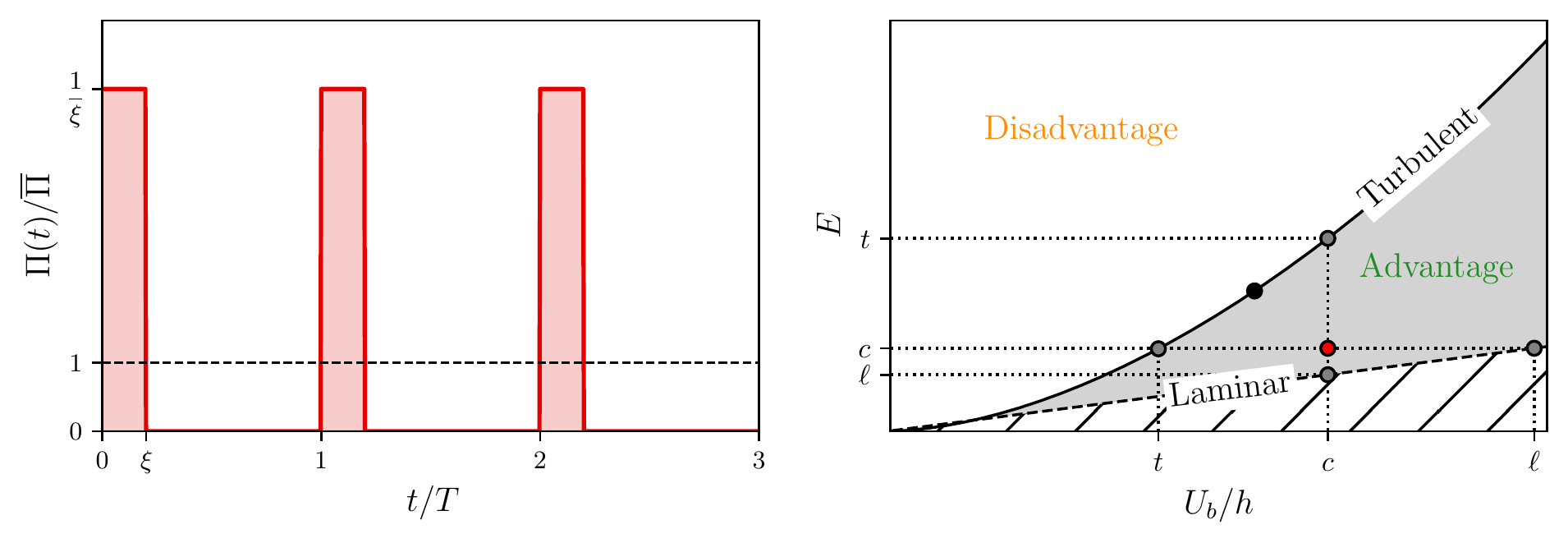}%
\caption{Left: temporal variation of the function $\Pi(t)$ employed in the present work, with period $T$ and duty cycle $\xi$. The cycle-averaged value $\overline{\Pi}$ is kept constant across the numerical experiments, as the maximum intensity is $\overline{\Pi}/\xi$. Right: sketch of the money-time plane, with the energy $E$ (money) needed to move the fluid through the duct on the vertical axis, and the inverse of the required time on the horizontal axis. The continuous line is the turbulent friction law, which describes all the uncontrolled flow states; the dashed line corresponds to the laminar regime. The black dot on the turbulent line represents a generic uncontrolled flow state, while the red dot, below the turbulent line, is a successfully controlled flow state, more efficient than the natural turbulent flow. The four grey dots on the turbulent and laminar lines are flow states used to quantify control performance (see text).} 
\label{fig:drawings}
\end{figure}

In the numerical simulation of a turbulent channel flow, a volumetric forcing is 
adopted to drive the fluid. It is customarily set to a constant value, or instead continuously adjusted to maintain a constant flow rate \citep{quadrio-frohnapfel-hasegawa-2016}.
In the present case, the forcing is a homogeneous, temporally periodic 
streamwise pressure gradient $-\Pi(t)$, whose waveform is a simple on-off pulsation 
with period $T$, switching between a prescribed constant value and zero, as 
schematically shown in the left panel of \cref{fig:drawings}.
Albeit inspired by the earlier work of \citet{iwamoto-sasou-kawamura-2007}, the present forcing does not lead to an adverse pressure gradient during the deceleration phase, and does not rely on energy recovery.

Whenever the pressure gradient is time-periodic, the number of parameters describing a parallel duct flow increases from one (typically, the Reynolds number) to three \citep{akhavan-kamm-shapiro-1991}: in this work, they are the period $T$, the duty cycle $\xi$ (i.e. the fraction of period with active pumping, $0 < \xi \le 1$), and the value of $\Pi$ during the \textit{on} phase.
The number of parameters reduces to two by constraining the cycle-averaged value $\overline{\Pi}$ of the pressure gradient to remain constant, as illustrated in \cref{fig:drawings} (left). In the present study, $\overline{\Pi}$ is set equal to the pressure gradient of a conventional turbulent channel flow at $Re_\tau \equiv u_\tau h / \nu = 180$.
Thus, the forcing term $\Pi(t)$ switches between $\Pi=\overline{\Pi}/\xi$ for a fraction $\xi T$ of the period, and $\Pi=0$ for the remaining part, i.e.
\[
\Pi(t) = 
\begin{cases}
\begin{aligned}
 \frac{\overline{\Pi}}{\xi} \qquad & \mbox{for} \quad nT \le t < (n+\xi) T , \\
 0                           \qquad & \mbox{for} \quad (n+\xi)T \le t < (n+1)T
\end{aligned}
\end{cases}
\]
where the integer $n$ indicates the generic $n$-th forcing cycle.

\subsection{The numerical setup}

Two different DNS solvers are used to ensure the robustness of the results. 
The first code \citep{mazzino-rosti-2021}, primarily used to produce all quantitative information, discretises the Navier--Stokes equations on a staggered, Cartesian grid, with a hyperbolic tangent distribution employed in the wall-normal direction.
The spatial derivatives are approximated using second-order finite difference schemes, while an explicit Adams--Bashforth scheme is adopted for the 
time-advancement, with a bound on the CFL number used to dynamically adjust the time-step. The pressure coupling is solved by implementing the fractional-step method proposed by \citet{kim-moin-1985}, with the resulting Poisson equation solved using a fast 
pressure solver. 

The second code \citep{luchini-quadrio-2006}, used to verify selected cases and to confirm the robustness of the most important claims, solves the normal-velocity normal-vorticity formulation of the Navier--Stokes equations as introduced by \cite{kim-moin-moser-1987}.
The equations are spatially discretised with a spectral method along the 
homogeneous directions, while fourth-order compact and explicit finite-difference schemes are adopted to approximate derivatives along the wall-normal direction.
A partially implicit approach, combining the explicit third-order low-storage 
Runge--Kutta and the implicit second-order Crank--Nicolson schemes, is used for the
temporal advancement of the equations, with the time step adjusted according to the 
CFL number imposed.

The present simulations involve pressure gradients and, consequently, flow rates 
that vary during the period $T$; this poses important challenges in terms of time and space discretisation. 
An extremely accurate time integration is needed, much like in stability and 
transition problems: the value of the CFL number, that typically assumes values at or above unity in turbulent channel flow simulations as long as the stability limit of the time integration scheme allows \citep{choi-moin-1994}, is kept at $0.1$ here, since it has been verified that higher values lead to spurious results affected by the time step size.

Several discretizations, in terms of size of the computational domain and spatial resolution, are considered. 
The latter is typically expressed in plus or viscous units; in the present, unsteady flow, a reference $Re$ must be selected to define viscous units. Since the chosen value of $\overline{\Pi}$ corresponds, for a steady forcing, to the value $Re_\tau=180$, the $+$ units employed throughout the paper are defined for $Re_\tau=180$.
The core of the study uses grid spacings of $\Delta x^+ = 6.6$ and $\Delta z^+ = 3.3$ along the homogeneous directions, with a minimum grid size of $\Delta y^+ = 0.5$ at the wall and a maximum of $\Delta y^+ =  3.2$ at the centreline.
Such sizes would be more than adequate for a conventional channel flow at 
$Re_\tau = 180$, but require further consideration here: in fact, during the pumping cycle, the instantaneous value of $Re_\tau$ grows by up to three times compared to the reference $Re_\tau=180$. Unfortunately, correspondingly finer meshes would lead to a computational load we cannot afford.
However, we have confirmed with a simulation on a coarser mesh (with $\Delta x^+ = 13.2$, $\Delta z^+ = 6.6$ and $\Delta y^+ = 0.6 - 4.1$) that the key quantity, i.e. the energy saving, computed for the best performing case, is essentially unchanged, with a relative 1\% variation only. 
Furthermore, a few forcing cycles have been recomputed with two successively refined grids, up to $\Delta x^+ = 2.2$, $\Delta z^+ = 1.1$, $\Delta y^+ = 0.09 - 1.2$. Although the number of cycles was insufficient for a proper statistical assessment, we have obtained the same temporal evolution for the key quantities described in figure \ref{fig:0D}.

Results are also sensitive to the size of the computational domain. All cases have been preliminarly run on a computational box with $L_x= 3 \pi h$, $L_y=2h$ and $L_z = 1.5 \pi h$, but all those providing drag reduction have been recomputed with the wall-parallel size of the computational domain increased to $6 \pi h \times 3 \pi h$, correspondingly incrementing the number of grid points. Two additional checks, with computational domains of $(L_x,L_z) = (9 \pi h \times 4.5 \pi h)$ and $(L_x,L_z) = (16 \pi h \times 3 \pi h)$, are discussed in \S\ref{sec:performance}.

The production simulations, starting from an initial flow field at $Re_\tau=180$, are run for nineteen cycles for cases leading to drag increase, and thirty-six for the others. In the cycle-averaging procedure the first cycle is always discarded, as it bears memory of the initial condition.

\subsection{A meaningful way to evaluate performance}
\label{sec:moneyversustime}

A proper metric is needed to evaluate the performance of the pumping strategy. 
The prescribed pressure gradient is constrained to always have a cycle-averaged 
value corresponding to that of a canonical turbulent channel flow at 
$Re_\tau=180$. It would thus be natural to compare the achieved flow rate with that at $Re_\tau=180$, and discover that the flow rate can be almost doubled.
However, there is an extra energy cost that needs to be factored in, so that an increased flow rate is not sufficient to define a successful control. Instead, it must be verified that the amount of energy required by the unsteady pumping gets used more efficiently than by the standard steady pumping.

A tool that is well suited for this assessment was introduced by \citet{frohnapfel-hasegawa-quadrio-2012}.
They represented in the so-called money-time plane the pumping energy $E$ 
required to move a given amount of fluid through a straight duct in a given time versus the quantity $h/U_b$ (with $U_b$ the bulk velocity), which represents the time taken by a fluid particle to travel for the reference length $h$, referring to the first variable as \textit{money} and to the second as \textit{time} or \textit{(in)convenience}. 
A natural turbulent flow is represented on this plane by a point which moves along a line (representing the turbulent friction law) as the Reynolds number is changed. 
The line can be drawn by e.g. resorting to the Blasius' correlation \citep{schlichting-1979} to link the friction coefficient $C_f$ with the Reynolds number, so that $C_f \propto U_b ^ {-1/4}$ and $E \propto U_b^{7/4}$. 
In \cref{fig:drawings} (right), the money-time plane is sketched (with the variable on the horizontal axis changed to $U_b/h$ with respect to the original version) with the addition of the laminar line, where $C_f \propto 1/U_b$ and $E \propto U_b$. 
The plane is partitioned in three regions. Theoretical arguments \citep{fukagata-sugiyama-kasagi-2009} ensure that the region below the laminar line is unreachable (even with active control, provided that the control energy is accounted for). 
The goal of successful flow control, which increases the energy efficiency of the flow, is to reach a point in the graph that sits in the grey area below the turbulent line (as the red dot in \cref{fig:drawings}). 
No unique way exists to move the natural flow state (black dot in \cref{fig:drawings}) towards the laminar line. 

To make the present analysis quantitative, a figure of merit is required. It can describe either energy savings, or improved performance, or both. 
Energy saving is expressed by the distance along the vertical axis between the controlled flow with energy expenditure $E_c$ and the point on the turbulent curve with the same flow rate and energy expenditure $E_t$. 
Thus, the energy figure of merit $S_e$ for the savings is $S_e = (E_t-E_c) / (E_t-E_\ell)$, where the denominator $E_t-E_\ell$ accounts for the impossibility to reach below the laminar curve, and thus represents the maximum possible savings. 
An analogous indicator that quantifies the improved convenience is defined by 
considering the achieved flow rate: 
$S_c= (1/U_{b,t} - 1/U_{b,c}) / (1/U_{b,t} - 1/U_{b,\ell})$. 

\section{Results}
\label{sec:results}

\subsection{Control performance}
\label{sec:performance}

The study samples the parameter plane ($T,\xi$) in twenty points with the finite difference code introduced in \S\ref{sec:setup}.
Eighteen points were determined after a previous, less resolved analysis \citep{monti-2015} carried out with the spectral code, where the periods $T^+\in[3600,10800,14400]$ and the duty cycles $\xi\in[0.005,0.0125,0.025,0.0375,0.05,0.1]$ were considered.
(Note again that, as stated above, viscous units are defined for a conventional channel flow at $Re_\tau=180$.)
Two additional points at the longer period $T^+=18000$ and $\xi\in[0.025,0.1]$ are added to complete the dataset.
The successful cases (namely, those with $T^+=10800$, $\xi\in[0.025,0.1]$ and all those with the longest periods $T^+=14400$ and $T^+=18000$) are run on the largest computational domain.

\begin{figure}
\centering
\includegraphics{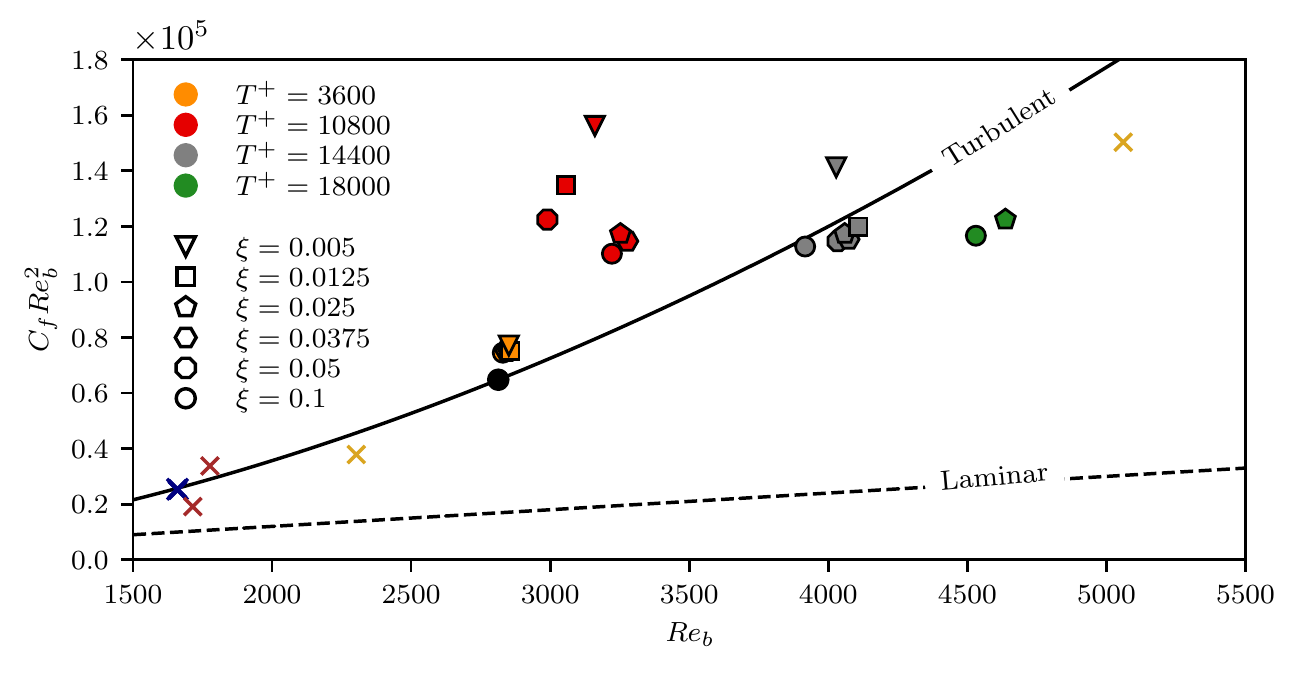}
\caption{DNS results (filled colour symbols) compared to the reference uncontrolled flow (continuous line), the laminar flow (dashed line) and relevant results taken from literature. 
The colour of the filled symbols encodes the forcing period, and their shape refers to the value of the duty-cycle. The results from \citet{iwamoto-sasou-kawamura-2007}, \citet{kobayashi-etal-2021} and opposition control data reported by \citet{frohnapfel-hasegawa-quadrio-2012} are represented as blue, red and yellow crosses, respectively. The black dot on the reference line corresponds to $Re_\tau=180$.} 
\label{fig:resMvsT}
\end{figure}

The outcome of the study is plotted in \cref{fig:resMvsT} on the money-time plane, where the two quantities on the coordinate axes are now made dimensionless with $Re_b$ (convenience) and $C_f Re_b^2$ (energy). The quantities $U_b$ and $C_f$ are time-averaged over the cycles, once the first cycle is discarded. Denoting the time average with an overbar, in particular, we compute $C_f=({2}/{\rho})({\overline{\Pi U_b}}/{\overline{U}_b^3})$.
The point corresponding to the canonical turbulent channel flow at $Re_\tau=180$ (or $Re_b \approx 2800$) is also plotted; by definition, it sits on the turbulent curve.
For comparison, we include results obtained by \citet{iwamoto-sasou-kawamura-2007} and the more recent ones presented by \citet{kobayashi-etal-2021}. 
Two points reported by \citet{frohnapfel-hasegawa-quadrio-2012} and corresponding to opposition control \citep{choi-moin-kim-1994} are highlighted as well.

Cases with the smallest period nearly overlap at various $\xi$; although their flow rate is increased slightly compared to $Re_\tau=180$, they all lie above the turbulent line and provide no savings. Cases at $T^+=10800$ are more scattered, but consistently produce negative performance. 
However, all the cases with the longest periods, i.e. $T^+=14400$ (with the exception of the smallest duty-cycle $\xi=0.005$) and $T^+=18000$ lay clearly below the turbulent curve, and attest the success of the control technique.
Although details are not shown here, this outcome is robust with respect to the number of periods considered for the averaging. We have measured that the value of the saving $S_e$ computed for the best-performing case by averaging over 35 cycles has reached statistical convergence, since its running average has a relative fluctuation of less than 3\% over the last periods of forcing. 


The best performing case ($T^+=18000$ and $\xi=0.1$) yields $S_e=0.27$ and $S_c=0.17$, i.e. a 27\% energy savings or, if preferred, a 17\% improvement in convenience. These figures can be compared with those by one of the most successful flow control approaches, namely the opposition control \citep{choi-moin-kim-1994}, designed to cancel turbulence in numerical experiments where the flow state is known, at all times and positions, in a sensing wall-parallel plane and a counteracting time-dependent distributed non-uniform blowing/suction is applied at the wall. From the work by \cite{frohnapfel-hasegawa-quadrio-2012}, the opposition control at a comparable Reynolds number (i.e., $Re_b \approx 5000$) yields $S_e \approx 0.23$ and $S_c \approx 0.13$. 
Note, however, that the opposition control is an idealized active control technique, requiring distributed real-time sensing within the flow, and real-time distributed actuation: its practical implementation is extremely difficult, while the present system is in principle simple, and requires neither sensors nor actuators. 

There is no reason to assume that the drag reduction made possible by unsteady pumping as measured in this study could not be outperformed. In fact, the parameter space and the temporal wave form have been only preliminarily explored in present study, whose results should only be considered as a lower limit for the drag reduction potential.
Furthermore, the streamwise length of the computational domain is not sufficient to provide a drag reduction figure that is truly domain-independent. 
Only for the best case at $T^+=18000$ and $\xi=0.1$, we have carried out two additional simulations in larger computational domains: their wall-parallel size is $9\pi \times 4.5\pi$ and $16\pi \times 3\pi$, discretised with grids of spacing $\Delta x^+ = 6.6, \Delta z^+ = 3.3, \Delta y^+ = 0.5 - 3.2$ and $\Delta x^+ = 13.2, \Delta z^+ = 6.6, \Delta y^+ = 0.6 - 4.1$, respectively. 
In both cases $S_e \approx 0.4$ has been reached, suggesting that a domain of wall-parallel size  $9\pi \times 3\pi$ might be needed to yield nearly domain-independent results. 

\subsection{Intra-cycle dynamics}

\begin{figure}
\centering
\includegraphics{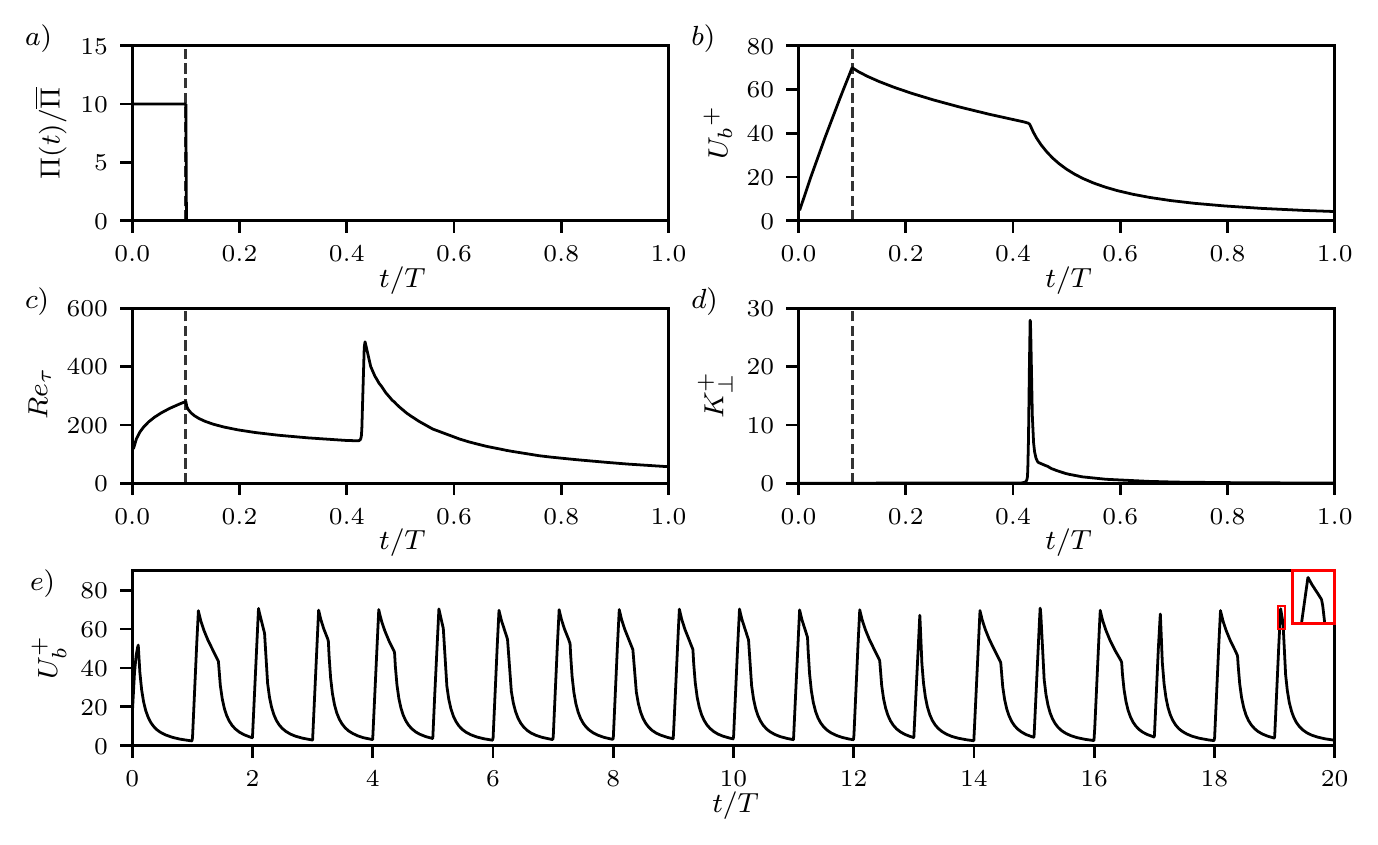}
\caption{Time evolution of the forcing term $\Pi$ (a), the corresponding flow rate (b), the friction-based Reynolds number $Re_\tau$ (c) and the cross-plane turbulent kinetic energy $K_\perp$ (d), during one representative cycle, for $T^+=14400$ and $\xi=0.1$. The bottom panel (e) shows the flow rate over the first 20 cycles, with the inset zooming in on the last peak.}
\label{fig:0D}
\end{figure}

The flow undergoes a significant evolution during each forcing cycle.  
The time evolution of the bulk velocity $U_b(t)$ is described by:
\begin{equation}
\frac{\partial U_b(t)}{\partial t}=\frac{\Pi(t)}{\rho}+\frac{\langle \tau_w \rangle(t)}{\rho h}
\label{eq:meanflow}
\end{equation}
where $t$ is the time, and $\langle \tau_w \rangle$ is the space-averaged wall shear stress. 
The balance above describes the interplay among flow inertia (whose time-average is zero), the driving pressure gradient and the wall friction at every instant of the forcing cycle.
In particular, during the acceleration phase the unsteady term balances the large pressure gradient while the wall shear stress builds up: a steady state would be eventually attained when ${\partial U_b}/{\partial t}=0$. 
During the deceleration, conversely, $\Pi$ is identically zero and the mean flow decays at a rate imposed by the wall shear stress.

Figure \ref{fig:0D} illustrates the flow dynamics within one representative cycle, in terms of the time evolution of the global quantities featured in the balance \eqref{eq:meanflow}.
As expected, when the driving pressure gradient $\Pi(t)$ is active in the first portion of the cycle (panel a), the flow is accelerated and the flow rate 
(i.e., the bulk velocity $U_b$) undergoes a significant growth (panel b). 
Although the flow rate would eventually settle to the new level consistent with the constant value $\overline{\Pi} / \xi$, on the time scale $\xi T$ the increase of $U_b$ is nearly linear, implying a strong non-equilibrium in which the pressure gradient overwhelms friction. During the acceleration phase, the streamwise velocity profile changes too, and its wall slope increases, as seen by the instantaneous value of $Re_\tau$ computed from the space-averaged and instantaneous wall stress ($Re_\tau=(h/\nu)\sqrt{\langle \tau_w \rangle / \rho}$) for $0 \le t \le \xi T$ (panel c). 
It is known \citep{greenblatt-moss-1999} that, during intense accelerations, turbulence can be destroyed, and the flow can approach re-laminarisation. 
Panel (d) plots the evolution of the volume-integrated cross-stream turbulent kinetic energy:
\[
K_\perp = \frac{1}{4h L_x L_z} \int_0^{2h} \int_0^{L_x} \int_0^{L_z} (v^2 + w^2) dx dy dz ,
\] 
which provides a direct indication of the turbulent state of the flow, while neglecting longitudinal velocity fluctuations around the time-varying streamwise velocity profile. This quantity is extremely small at the beginning of the acceleration, and remains so for a rather long time after $t/T=\xi$.  
Once the pumping is turned off at $t/T=\xi$, the flow slows down because of viscous losses: $U_b$ and $Re_\tau$ decrease, while $K_\perp$ remains nearly zero. However, after some finite time delay (at about $t/T \approx 0.45$ for the period considered), the decay rate of $U_b$ suddenly increases, and a noticeable kink is observed in the $U_b(t)$ curve of panel (b). 
At the same time instant, $Re_\tau$ peaks again, reaching the largest value during the cycle, and $K_\perp$ quickly rises to a large local maximum: in a very short time, the flow becomes turbulent. 
Note that, the later the kink appears in the curve of the flow rate, the larger is the cycle-averaged flow rate, and thus the drag reduction, as in this part of the cycle pumping is off, and a larger flow rate comes at no extra energy costs. 
After this sudden transition to turbulence, in the absence of pumping the flow decays normally until, at the end of the cycle, $U_b$ is very low, $Re_\tau$ reaches a small value, and $K_\perp$ becomes nearly zero again. The remnants of turbulence are further annihilated during the following acceleration; however, traces of the turbulent structures remain which eventually determine the next transition during the following pumping cycle.

The lower panel of figure \ref{fig:0D} shows a fraction (twenty periods) of the whole time history for the flow rate, including the first cycle that is directly affected by the initial condition and is thus discarded when computing statistics. 
The qualitative behaviour described above in panel (b) is observed at every cycle, but with significant quantitative inter-cycle variations.
The kink in the curve of the flow rate which marks the onset of turbulence is always present, but the time of its occurrence changes, ranging from very shortly after the pump being switched off (as in the last period reported in the red inset) to the middle of the forcing cycle (as in the second period).
The time delay between the end of the acceleration and the kink should therefore be regarded as a random variable, whose mean value affects drag reduction.
The time history also shows an apparently random appearance of subharmonic dynamics. An example is the two-periods pattern, visible from $t/T=12$ onwards in panel (e) of figure \ref{fig:0D}, in which the kink appears alternately very shortly after the end of acceleration, and rather far from it.  
Visual analysis of the whole set of time histories reveals that such events involve a variable number of periods and appear randomly in time, possibily due to a lock-in between the forcing and a characteristic frequency of the system.
While this analysis is beyond the aim of this preliminary study, we cannot rule out the possibility that such dynamics has a role in determining the overall drag reduction.



\subsection{Effect of the control parameters}

To asses the effect of the control parameters ($T$ and $\xi$), it is instructive to compare across all cases two key quantities extracted from each pumping cycle and then averaged together, namely the maximum value of the bulk velocity and the time interval that separates this maximum and the transition to turbulence.
The former quantity, obtained by averaging the values of the bulk velocity at the end of the pumping phase, is
\[
U_{b,m} = \frac{1}{N}\sum_{i=1}^N \left( \max_{iT\leq t< (i+1)T}{(U_b(t))} \right) ,
\]
where $N$ is the total number of periods after discarding the first. 
The latter quantity, marking the average extent of the quasi-laminar flow phase extending between the end of the pumping and the transition to turbulence, is defined as the time delay $\tau$ between the end of acceleration, occurring for cycle $i$ at the time $(i+\xi)T$, and the time $t_K^{(i)}$ where the peak of the cross-stream turbulent kinetic energy $K_\perp$ occurs:
\[
\tau = \frac{1}{N} \sum_{i=1}^N \left( t_K^{(i)} - (i + \xi) T \right) .
\]
In particular, $U_{b,m}$ is related to the energy spent to accelerate the fluid, while $\tau$ determines the energy savings. 
\begin{figure}
\centering
\includegraphics{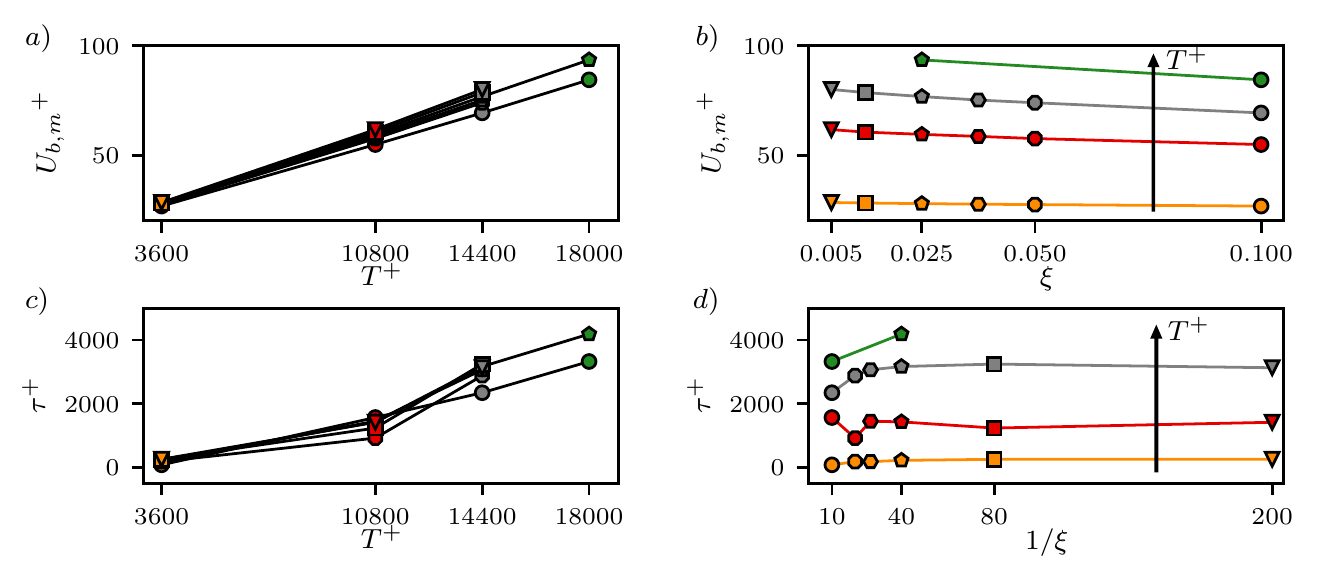}
\caption{$U_{b,m}^+$ (top row) and $\tau^+$ (bottom row) against $T^+$ (left column) and $\xi$ (right column). Symbols and colors as in figure \ref{fig:resMvsT}.}
\label{fig:scaling}
\end{figure}
When plotted against $T^+$ (in figure \ref{fig:scaling}, left), both quantities collapse well (albeit not perfectly) for various values of the duty cycle. The collapse of $U_{b,m}$ (panel a) would be perfect if the growth of $U_b$ was linear and proportional to the maximum pressure gradient $\overline{\Pi} / \xi$. 
It is thus not surprising that cases where the acceleration lasts longer (at $\xi=0.1$ and large periods) are those showing less overlap, as there is enough time to see at least the beginning of saturation in the growth of $U_b$. 
For the same reason, cases with large periods do not produce a perfectly horizontal line in panel (b) when plotted against $\xi$.
The outcome in terms of $\tau$ observed in panel (c) is less predictable and perhaps more interesting: its value is minimal for small $T$, but seems to increase non-linearly at higher $T$. 

In terms of control performance, a large delay $\tau$ is always beneficial: the flow remains in a quasi-laminar state with a large $U_b$ for a longer fraction of the cycle, while no pressure gradient and thus no energy expenditure is required to produce the flow rate. 
Regardless of $\xi$, increasing $T$ always yields a larger $\tau$ (panel c) and net savings (i.e. overcoming the increased energy cost due to the larger $U_{b,m}$, panel a).
When instead $T$ is fixed, $U_{b,m}$ decreases with $\xi$ (panel b) while $\tau^+$ exhibits a plateau for the lowest values of $\xi$ (panel d).
It is therefore to be expected that, with the present forcing, the largest savings are found on the plateau (where $\tau^+$ is almost constant) for the lowest $U_{b,m}$.

\section{Conclusions}
\label{sec:conclusions}

We have observed by DNS of turbulent channel flows that a flow control strategy which periodically modulates in time the pumping power can improve the energetic efficiency of the pumping process above that of a steady pumping. 
The modulation considered in the present work is a simple sequence of on/off pulses, described by the duration of the cycle and the duration of the \textit{on} phase, with its amplitude varying as the inverse of the duty cycle. Alternative waveforms are easily conceivable, and may lead to even better performance. The control approach enjoys some practical appeal, as there is no need for sensors and/or actuators.

A rational criterion is needed to ascertain the success of the flow control strategy. Our approach avoids the misleading comparison with a predetermined (arbitrary) reference flow, and only verifies whether the energy spent has been actually spent better than for a steady pumping. 
The achieved benefits are already higher than many active and passive flow control techniques documented in the literature, and result from a simple approach.
Moreover, in the present study only few numerical experiments are carried out, computational limitations may underestimate the benefits, and only one temporal waveform is considered: a huge parameter space remains to be explored. 
However, it can be already claimed that on-off pumping can be successful, and that a steady power injection is not necessarily optimal from an energetic viewpoint.

Energy savings are obtained when the cycle period is large, leading to large flow rates at the end of the acceleration phase. 
We have observed that savings are obtained when the flow spends a significant fraction of the cycle in a transient quasi-laminar state, in which pumping has been just turned off but the flow rate decays slowly thanks to the absence of turbulent activity. 
The quantitative features of this peculiar flow state (in particular its duration) have been verified to be robust with respect to the discretisation schemes and spatial resolution. The cycle-averaged duration of the quasi-laminar state grows with the forcing period, which is the main reason why long pulsations are necessary to obtain drag reduction.

In terms of practical applications, further steps are needed to verify the feasibility of this strategy: e.g., thermal currents or background noise might alter the cyclic transition between the laminar and the turbulent states. 
The role played by the finite domain and by the periodic boundary conditions should be carefully evaluated. 
Furthermore, the alternate power surges needed to drive the system might constitute a significant practical drawback.
Nevertheless, the most interesting follow-up, that we leave for a forthcoming paper, is of a more fundamental nature, and concerns the detailed study of the peculiar laminarisation and transition processes that take place during the pumping cycle. 

\backsection[Acknowledgements]{M.E.R. acknowledges the support of the Okinawa Institute of Science and Technology Graduate University (OIST) with subsidy funding from the Cabinet Office, Government of Japan. The authors acknowledge the computer time provided by the Scientific Computing section of the Research Support Division at OIST, and useful discussion with Dr.-Ing. Davide Gatti at KIT.}

\backsection[Declaration of interests]{The authors report no conflict of interest.}

\backsection[Author ORCIDs]{

G. Foggi Rota, https://orcid.org/0000-0002-4361-6521; 

A. Monti, https://orcid.org/0000-0003-2231-2796; 

M. E. Rosti,  https://orcid.org/0000-0002-9004-2292; 

M. Quadrio, https://orcid.org/0000-0002-7662-3576}

\bibliographystyle{jfm}

\end{document}